# Use Non-Energy-Curtailment Resources for Primary Frequency Response in Future Low-Inertia Power Grids

Shutang You

*Abstract*— **Power grid primary frequency response will be significantly impaired by Photovoltaic (PV) penetration increase because of the decrease in inertia and governor response. PV inertia and governor emulation requires reserving PV output and leads to solar energy waste. This paper exploits current grid resources and explores energy storage for primary frequency response under high PV penetration at the interconnection level. Based on the actual models of the U.S. Eastern Interconnection grid and the Texas grid, effects of multiple factors associated with primary frequency response, including the governor ratio, governor deadband, droop rate, and fast load response, are assessed under high PV penetration scenarios. In addition, performance of batteries and supercapacitors using different control strategies is studied in the two interconnections. The paper quantifies the potential of various resources to improve interconnection-level primary frequency response under high PV penetration without curtailing solar output.**

## I. Introduction

Photovoltaic (PV) and wind generation are increasing in many power grids. The increase of PV penetration has an inevitable impact on the power grid primary frequency response due to the de-commitment of conventional units and consequent loss of inertia. As PV is usually set to work at the Maximum Power Point Tracking mode, it can hardly provide primary frequency response [1]. If PV is required to provide primary frequency response, some headroom needs to be reserved and a trade-off between the reserve capacity and the potential costs should be involved for economic considerations.

In spite of the trend of reducing conventional generation units in many power grids, many synchronous generators are still not likely to be retired and certain types of conventional generation may even expand in the foreseeable future. For example, hydro and nuclear generation units are more reliable than most renewable generation units. In addition, some power grids are expanding flexible thermal units powered by cheap and relatively clean fuels, such as natural gas [2]. These units have the potential to provide primary frequency response. Under the current grid operation paradigm, the frequency response capabilities of these synchronous units will still play an important role in providing primary frequency response to maintain frequency stability under high renewable penetration. The power industry has long observed that a large margin exists for implementing stricter frequency response settings [3]. Considering primary frequency response is an intrinsic capability of these synchronous units, it is more reasonable to leverage existing resources to the largest feasible extent than reserving headroom from renewable generation. However, few studies have exploited the primary frequency response capabilities of various existing resources in high renewable systems at the interconnection level.

With declining prices, energy storage systems are increasingly participating in short-term load balancing and peak load shifting [4]. Energy storage systems can be categorized into two groups in terms of their discharge time: high-energy-density energy storage and high-power-density energy storage. As a typical high-energy-density energy storage, chemical batteries have been used for primary frequency response in microgrids and small systems. Compared with batteries, electric double-layer capacitors (supercapacitors) have higher power-density but shorter discharging time. In addition, they have higher efficiency, shorter charge time, lower equivalent series resistance, and virtually unlimited cycle life. Compared with conventional capacitors, supercapacitors have higher energy capacities. Some studies have started to use supercapacitors to provide short-term power pulse for various applications, including smoothing fast power variations of wind [5, 6] and PV generation [7, 8], as well as participating in microgrid control [9]. With typical discharge time varying from 0.3 to 30 seconds and the capability of charging/discharging for millions of times, supercapacitors are ideal short-term energy sources for primary frequency response. Nevertheless, few studies have studied applying supercapacitors in supplying interconnection primary frequency response.

| Category | Tactics/Resources | | No. |
|---|---|---|---|
| Generation | PV and WTG inertia and governor emulation | | RE |
| | Synchronous units | Governor droop | SG1 |
| | | Governor db. | SG2 |
| | | Governor ratio | SG3 |
| Load | Fast responsive load | | FRL |
| Energy storage | High-energy-density storage | | ES1 |
| | High-power-density storage | | ES2 |

Fig. 1. Potential resources for primary frequency response at the interconnection level

This study offers a comprehensive understanding of the potential of interconnection-level frequency response resources under high PV penetrations. Two actual interconnection grids in the U.S. are used as the study systems: the Eastern Interconnection (EI) and the Electric Reliability Council of Texas (ERCOT) system. Based on detailed dynamic models of the two systems, various tactics for primary frequency response are studied, including leveraging resources at both the synchronous generation side and the load side, as well as batteries and supercapacitor energy storage. A summary of all potential resources for primary frequency response at the interconnection level investigated in this study is given in Fig. 1. Each resource is explored according to engineering feasibility and its current setting in each interconnection to quantify its potential to improve primary frequency response. Batteries and supercapacitors using droop frequency control and step response control are implemented in two interconnection grids. Additionally, this study reveals the impacts of discharge duration and energy storage capacity on primary frequency response metrics.

## II. Background: Primary Frequency Response and High PV Penetration EI and ERCOT Models

### *A. Key Metrics in Primary Frequency Response*

Primary frequency response evaluates the system frequency stability after major disturbances. It is essential to ensure the reliability of a power grid. For actual large-scale interconnection-level systems, frequencies may vary at different locations due to inter-area oscillations. To assess the interconnection-level frequency, the frequency value in this study is calculated using the central of inertia (COI) frequency:

$$f_{sys} = \frac{\sum_{i=1}^{N} H_i \cdot f_i}{\sum_{i=1}^{N} H_i} = \frac{1}{H_{sys}} \sum_{i=1}^{N} H_i \cdot f_i \tag{1}$$

where $H_i$ and $f_i$ are respectively the inertia constant and the frequency of the *i*th synchronous generator. Metrics for frequency response assessment using the COI frequency are shown in Table 1.

Table 1. Metrics in primary frequency response assessment [10]

| Metric | Definition |
| --- | --- |
| Nadir frequency ( $f_C$ ) | The minimum frequency after a resource contingency. This point is usually called 'C' point |
| Nadir time ( $t_C$ ) | Time to reach the nadir point after a contingency |
| Under frequency load shedding time ( $t_{UFLS}$ ) | Time to cross the under-frequency load shedding (UFLS) threshold after a contingency |
| Rate of change of frequency (ROCOF) | Frequency decline slope during the first 0.5s after a contingency |
| Frequency response ( $FR$ ) | Power imbalance over the frequency deviation (from the starting frequency to the settling frequency) |
| Nadir-based frequency response ( $FR_N$ ) | Power imbalance over the frequency deviation from the starting frequency to the frequency nadir |

### *B. High PV EI and ERCOT Models*

The geographic locations of the EI and ERCOT system in the U.S. are shown in Fig. 2. The two systems are modeled using positive sequence dynamic models in PSS/e®. For the EI system, the multi-regional modeling working group (MMWG) of Eastern Interconnection reliability assessment group (ERAG) built and maintains the power flow and dynamic model, which includes 68,309 buses and 8,337 generators. The EI MMWG model was developed by consolidating the regional dynamic models, whose component model types and parameters were collected from generation and transmission owners. This full-detail MMWG model is used as the base model for the EI system study. The ERCOT model is also an industry-owned model, consisting of 6,102 buses, 690 of which are generator buses. Statistic data of the base models for the two systems are shown in Table 2. The network diagrams of the U.S. EI and ERCOT systems are shown in Fig. 3. A summary of generation equipment dynamic model types used in this study is shown in Table 3. The frequency responses of both systems' base models have been validated using synchronphasor measurements from FNET/GridEye (a wide-area synchronphasor measurement system) and frequency events confirmed by utility companies. Model validation examples for each interconnection grid are shown in Fig. 4.

To reflect the industry outlook needs and maximize the study benefits, a survey was conducted within this study's technical review committee (TRC), whose service areas cover most EI and ERCOT regions, to determine high PV penetration scenarios of interest. According to the survey results, four simulation scenarios were eventually determined by TRC members as presented in Table 4. In both interconnections, PV penetration increases from 5% to 65% while wind penetration stays at 15% to study the PV impact. For each penetration scenario of each system, the geographic distribution of PV generation was projected using PLEXOS [11]. As an example, the PV distribution in the 80% renewable scenario is shown in Fig. 2. It can be noticed that regions in the southern and northeastern EI (close to load centers) have higher PV penetration rates. This is understandable since the southern EI has higher solar irradiance and the northeastern EI has higher local marginal prices and thus more economic surplus for PV. In the ERCOT, PV is more distributed in the west ERCOT due to high solar irradiance. The PV and wind generators used the General Electric PV and GE wind generation dynamic models for grid studies [12, 13]. These renewable generation dynamic models are average-value models suitable for large-scale power grid impact studies. According to the projected PV and wind power distribution in each scenario, the PV and wind models were incorporated into the base case model.

The largest resource contingencies in the EI and ERCOT are 4.5 GW and 2.75 GW generation loss, respectively, as specified by the North American Electric Reliability Corporation (NERC) in the Resource Contingency Criteria (RCC) [10]. Under different PV penetration levels, the EI and ERCOT system frequency responses after RCC contingencies are shown in Fig. 5 and Fig. 6, respectively. It shows that the frequency response deteriorates as PV penetration increases for both systems. Moreover, for the ERCOT, if no mitigation tactic (including fast load response) is applied, its frequency may cross the under-frequency load shedding threshold and cause load shedding when the renewable penetration rate passes 40%.

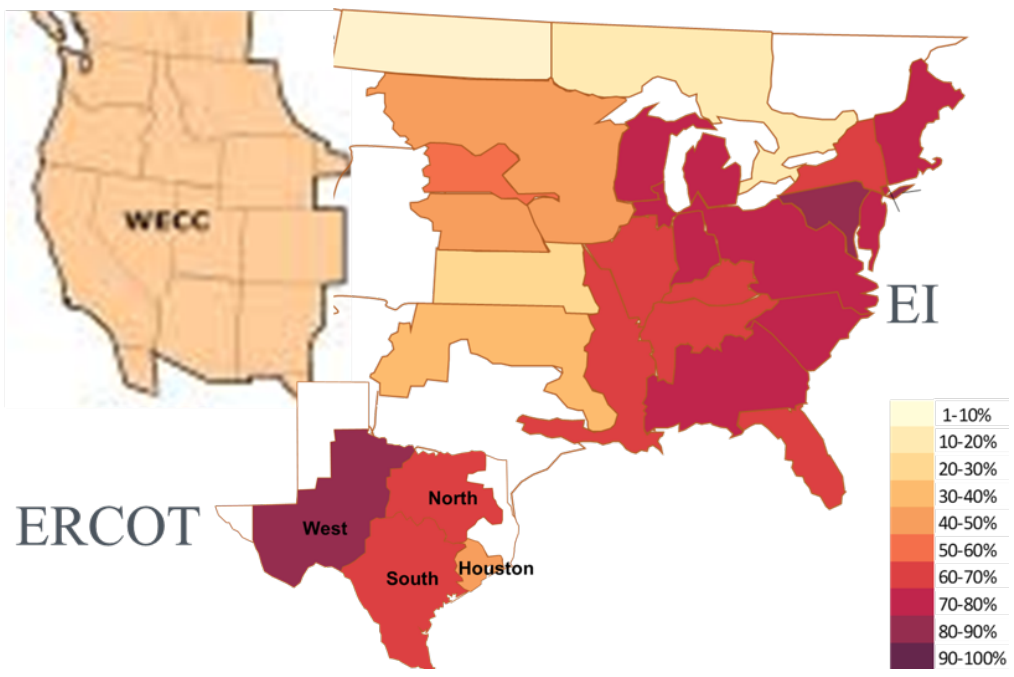

Fig. 2. U.S. EI and ERCOT locations in the U.S. and PV geographic distributions in the 80% renewable scenario

Table 2. EI and ERCOT model information

| System | Statistical metric | Value |
|---|---|---|
| EI | Total load | 560GW |
| | Total number of buses | 68,309 |
| | Total number of branches | 58,784 (non-transformer) + 21,460 (transformer) |
| | Total number of generators | 8,337 |
| | Voltage levels modeled | 0.69kV - 750kV |
| | RCC contingency magnitude | 4.5GW |
| ERCOT | Total load | 75GW |
| | Total number of buses | 6,102 |
| | Total number of branches | 6,319 (non-transformer) + 1,050 (transformer) |
| | Total number of generators | 690 |
| | Voltage levels modeled | 2.2kV - 350kV |
| | RCC contingency magnitude | 2.75GW |

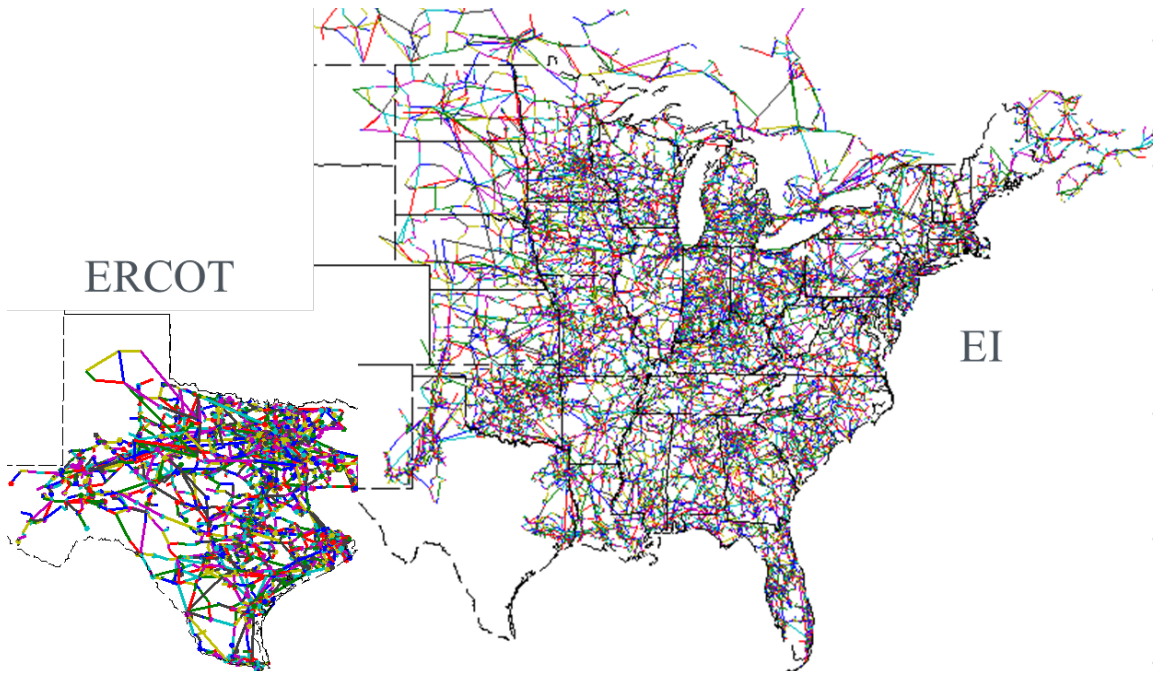

Fig. 3. Network diagrams of the U.S. EI and ERCOT system models

Table 3. Types of PSS/e dynamic models used in this study

| Devices | Models (in PSS/e®) |
|---|---|
| Synchronous generator | GENROU, GENCLS, GENSAL |
| Exciter | ESAC1A, ESDC1A, ESST1A, EXAC1, EXPIC1, EXST1, etc. |
| Governor | GAST, TGOV1, URGS3T, WSIEG1 |
| Stabilizer | PSS2A, PSS2B, IEEEST, STAB1-4, etc. |
| PV generation [12] | GEPVG (inverter), GEPVE (electric control) |
| Wind generation (Type 3 WTG) [13] | GEWTG2 (generator), GEWTE2 (electric control), GEWTT1 (wind turbine control) |

Table 4. Renewable penetration of all studied scenarios in the EI and ERCOT

| Scenario | Instantaneous PV Penetration | Instantaneous WTG Penetration | Total Renewable Penetration |
|---|---|---|---|
| # 1 | 5% | 15% | 20% |
| # 2 | 25% | 15% | 40% |
| # 3 | 45% | 15% | 60% |
| # 4 | 65% | 15% | 80% |

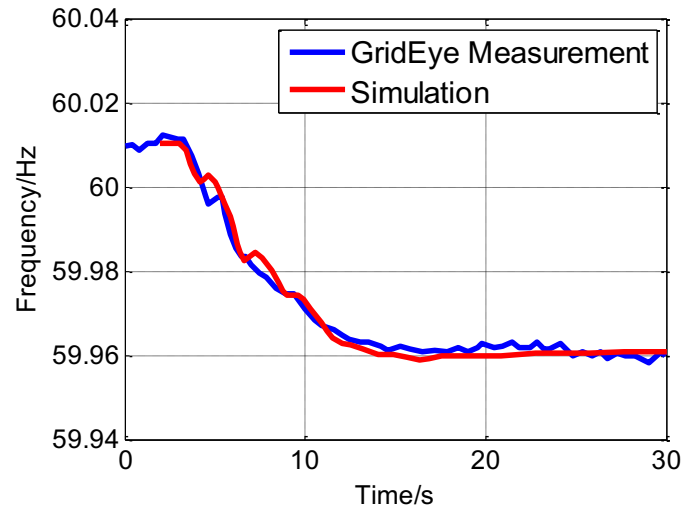

(a) EI

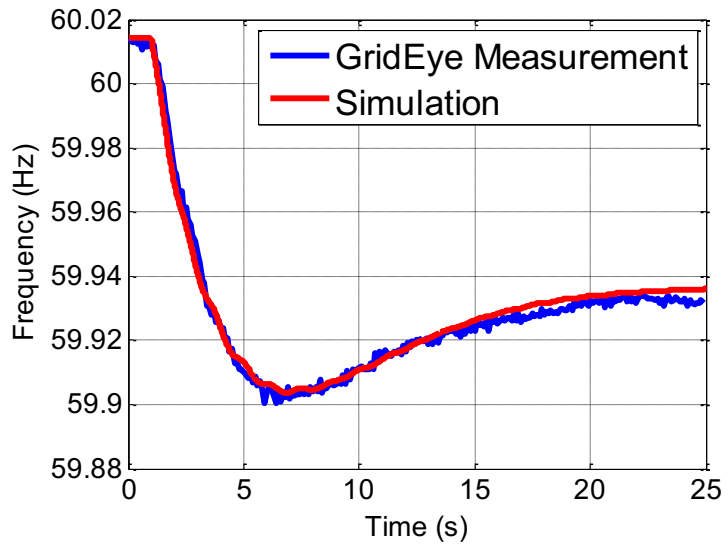

(b) ERCOT

Fig. 4. Examples of model validation result using synchrophasor data

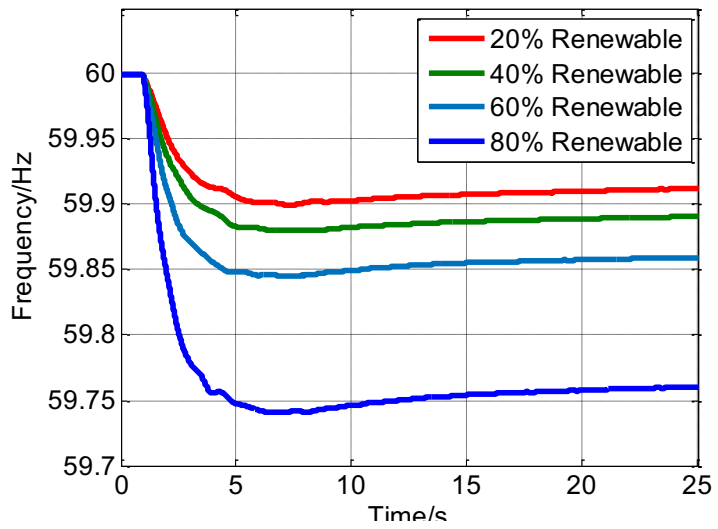

Fig. 5. EI frequency response under different PV penetration levels

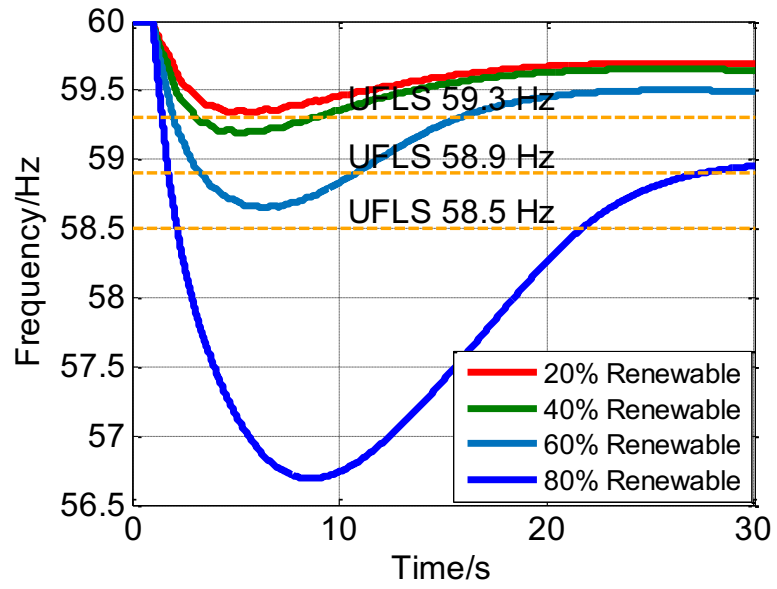

Fig. 6. ERCOT frequency response under different PV penetration levels

## III. EI System Primary Frequency Response Enhancement Without Curtailing Solar Output

### *A. Tactic SG1 — Synchronous Generation: Decrease Governor Droop*

The governor droop rates[1] of synchronous generators determine governors' steady-state reaction to frequency deviations . To investigate its potential in improving primary frequency response, the governor droop rates of all synchronous generators in the EI were reduced from 5% to 3% (the lower boundary recommended by the Western Electricity Coordinating Council [14]). As shown in Fig. 7 (only the 20% and 80% renewable cases are presented due to the page limit), 3% governor droop can improve frequency nadir and settling frequency significantly. This adjustment does not require additional governor equipment and solar energy curtailment. Nevertheless, detailed validation should be performed on actual system models before applying lower than 3% governor droop, since very small droop may lead to low frequency oscillations and other system instability issues.

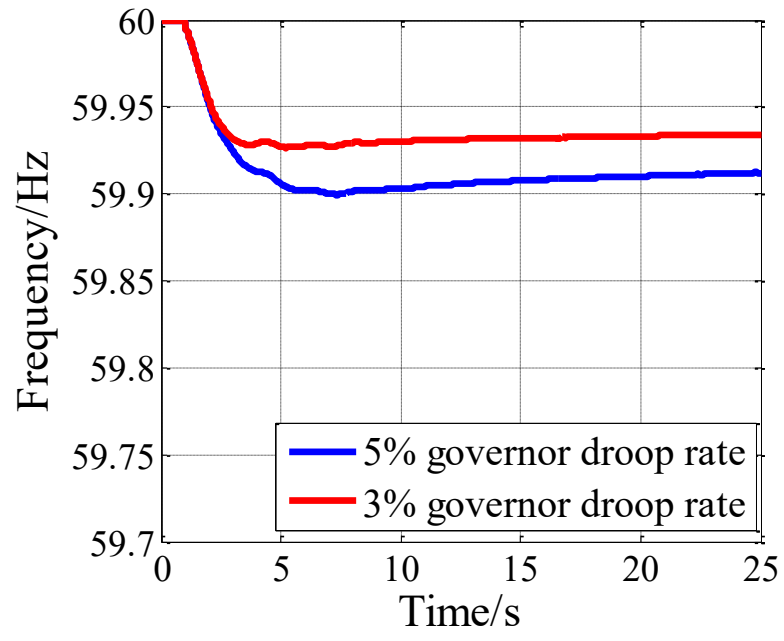

(a) 20% renewable

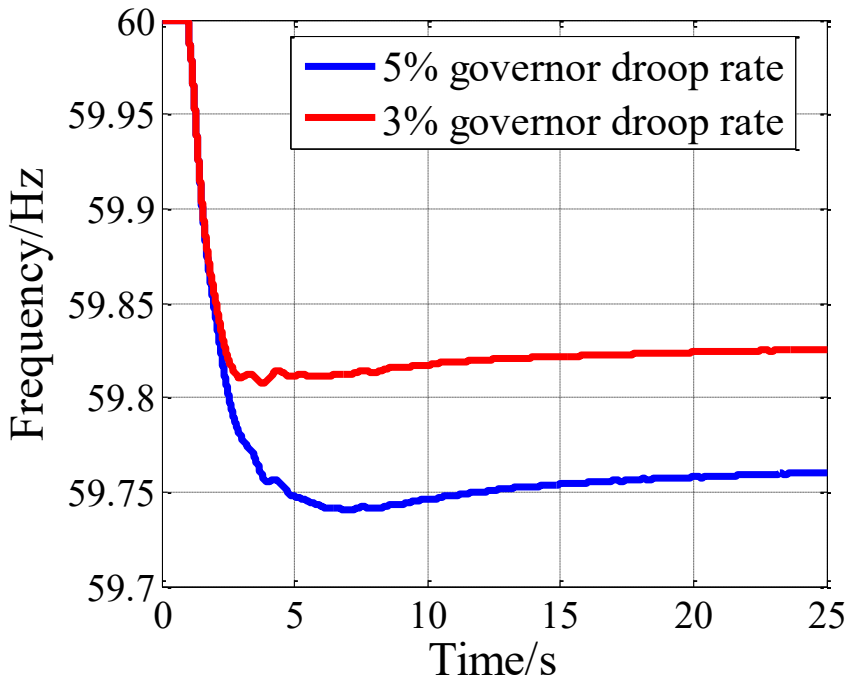

(b) 80% renewable

Fig. 7. EI frequency responses with different governor droop settings (4.5 GW generation loss in the RCC contingency)

### *B. Tactic SG2 — Synchronous Generation: Reduce Governor Deadband*

Governor deadbands commonly exist in actual power grids to prevent excessive governor reactions to small and frequent load variations. The existence of deadbands may delay governor responses to contingencies. Decreasing governor deadband could improve primary frequency response. Under high PV penetration, the average governor deadband in EI was narrowed down from 36 mHz to 16.7 mHz (NERC-recommended value for ERCOT [15]). Applying the RCC contingency, simulation results with two different governor deadband settings in 20% and 80% renewable penetration EI are shown in Fig. 8. It can be seen that decreasing governor deadband improves the EI frequency nadir and settling frequency. However, the improvement magnitude of settling frequency is a fixed value, which equals to the deadband reduction magnitude. Considering that common deadband values range from 16.7 mHz to 36 mHz but the frequency deviation may be much larger than this range under high PV penetration, the effectiveness of decreasing governor deadband is limited to small frequency deviation conditions (low renewable penetration and small contingency magnitudes).

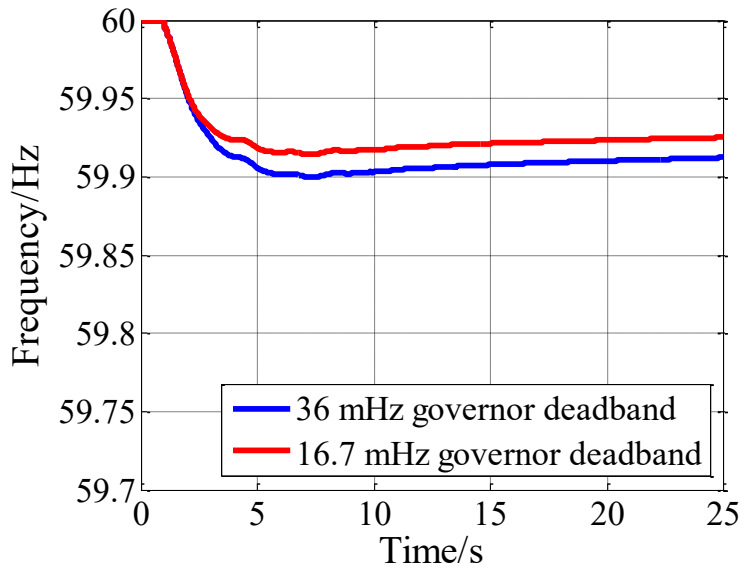

(a) 20% renewable

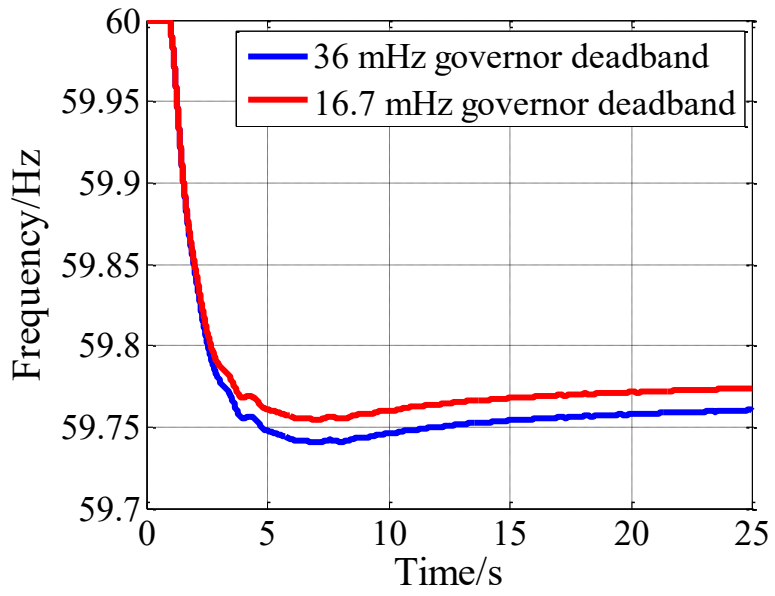

(b) 80% renewable

Fig. 8. EI frequency responses with different governor deadbands (4.5 GW generation loss in the RCC contingency)

### *C. Tactic SG3 — Synchronous Generation: Increase Governor Ratio*

According to some utility companies' statistics and study findings, as well as model validation experience using FNET/GridEye measurement, around 80% generation in EI can provide governor response but only around 30% governor resources are actually in service [16, 17]. To prevent frequency response from declining, those "idle" governors can be put into service in the future. Therefore, governor ratios were adjusted and its impacts on the EI frequency response were studied in the RCC contingency. As shown in Fig. 9, both frequency nadir and settling frequency were improved significantly as more governors contributed to primary frequency response. This result shows that increasing governor ratio is effective in improving system frequency response for a range of PV penetration scenarios. To encourage generators to participate in governor response, an ancillary service market may be necessary to procure adequate governor response for primary

[1] Droop is usually expressed as the percentage change in turbine/generator speed required for 100% governor action. For example, with 5% droop the full-load speed is 100% and the no-load speed is around 105%.

frequency response.

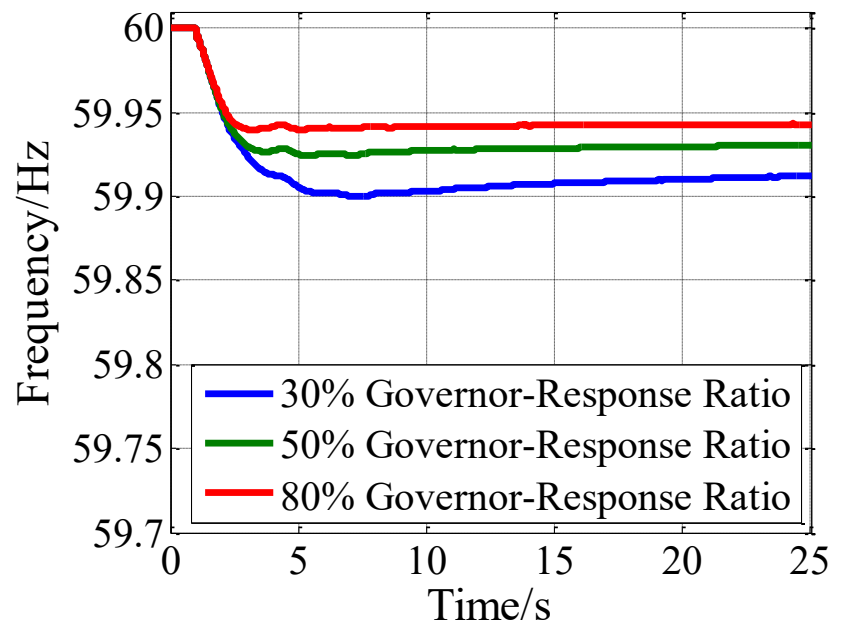

(a) 20% renewable

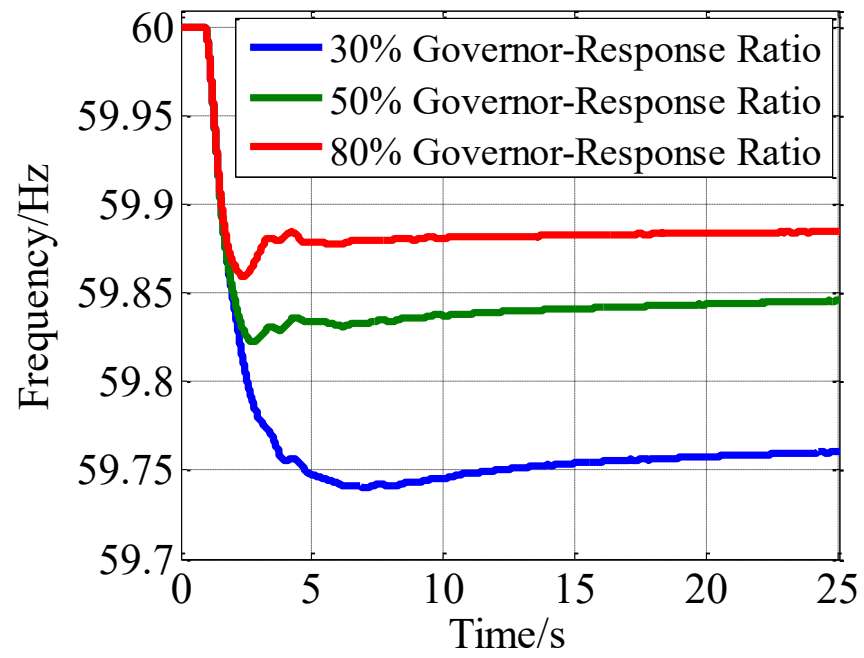

(b) 80% renewable

Fig. 9. EI frequency responses with different governor ratios (4.5 GW generation loss in the RCC contingency)

## *D. Tactic ES1 — Energy Storage: Battery*

Energy storage systems are increasingly used in PV systems to smooth output and balancing the load on a daily basis. Batteries and supercapacitors, as typical energy storage systems, can also provide primary frequency response using an ancillary frequency controller. Fig. 10 shows a high-level connectivity of the energy storage model and the frequency control model used for both batteries and supercapacitors. In this control model, frequency and voltage are measured at the point of grid-connection and fed back to the energy storage power controller to control real power output and terminal voltage. Parameters of the frequency control using batteries and supercapacitors are shown in Table 5.

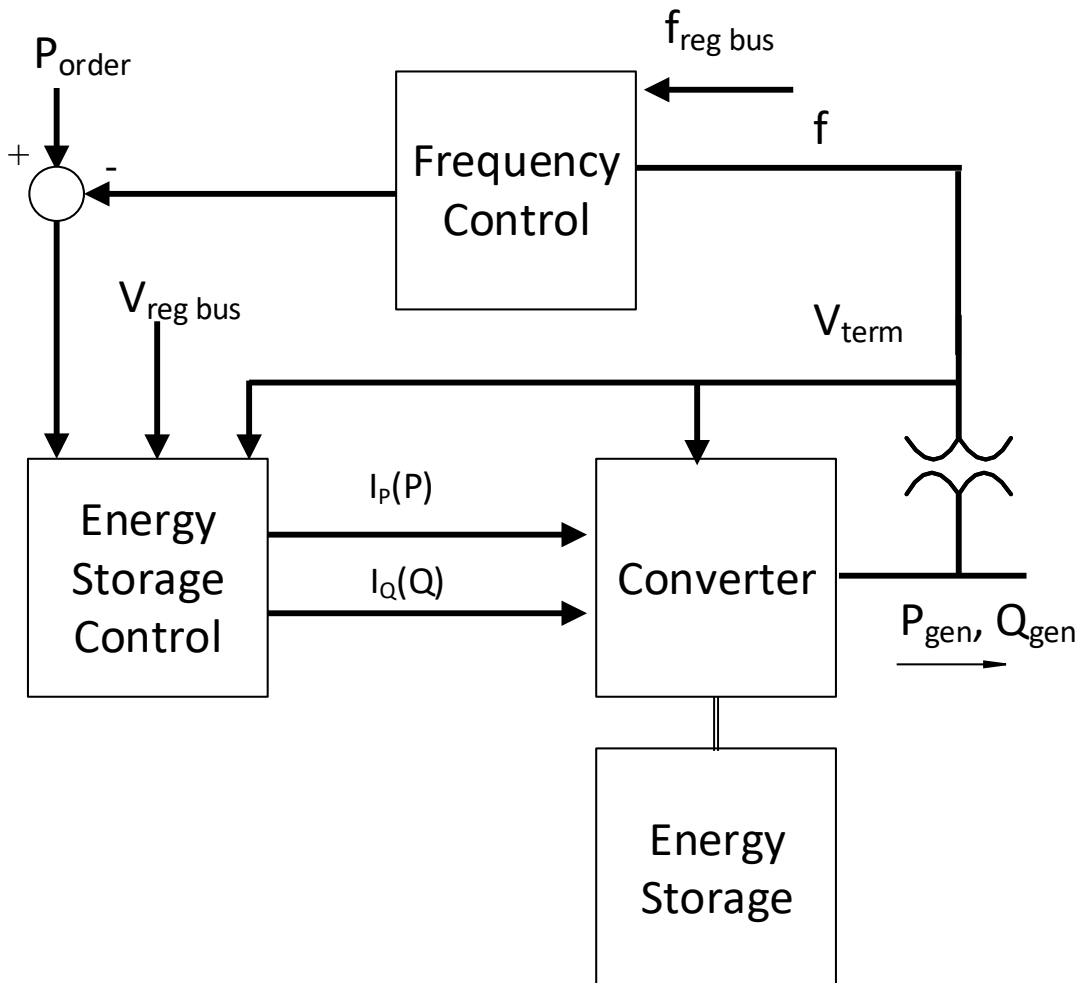

Fig. 10. Energy storage primary frequency control model

Table 5. Parameters of energy storage primary frequency control in the EI

| Parameter | Value |
|---|---|
| Energy storage maximum output | 3,100 MW |
| Number of energy storage installation locations | 100 |
| Droop frequency control droop rate | 3% |
| Droop frequency control deadband | 17 mHz |
| Step response threshold | 59.85 Hz |
| Step response delay for fault ride through | 0.5 s |
| Step response ratio α | 0.85 |
| Energy limit (default value for supercapacitors) | 3,100 MW*5s |

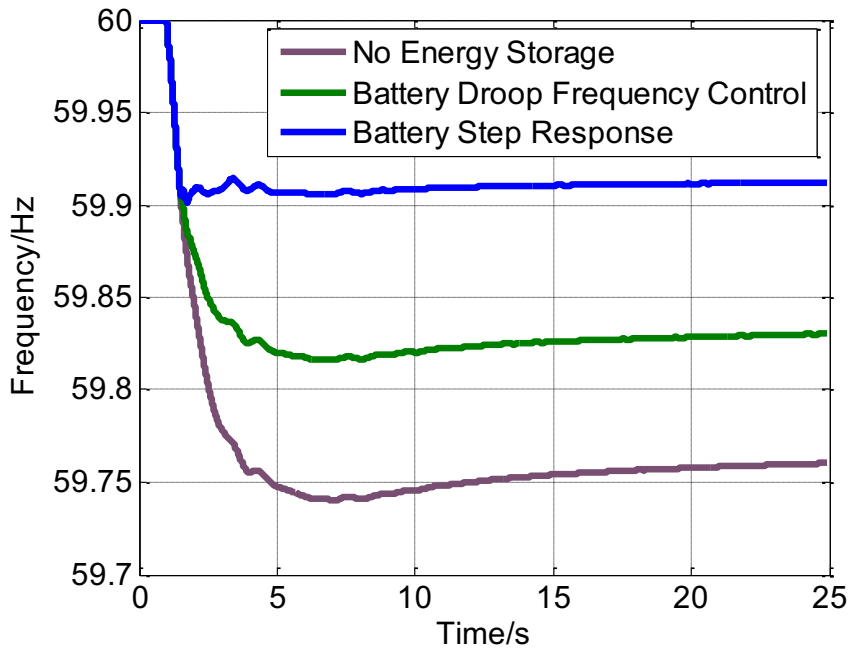

Fig. 11. EI frequency responses using droop frequency control and step response control of batteries (4.5 GW loss, 80% renewable)

Since batteries' typical discharge time ranges from 1-10 hours, it is reasonable to assume that stored energy in batteries is sufficient to provide sustained support over the time horizon of primary frequency response. The maximum instantaneous output of battery energy storage is constrained by the converter current limit.

Two typical control strategies are studied for energy storage frequency response: droop frequency control and step response control.

1) Droop frequency control: The droop frequency control mimics governor frequency regulation of synchronous generators. The measured frequency is compared with the reference frequency. The deviation passes a low-pass filter and then a control gain link to generate a power order signal as the input to the inverter controller. The control diagram is shown in Fig. 12.

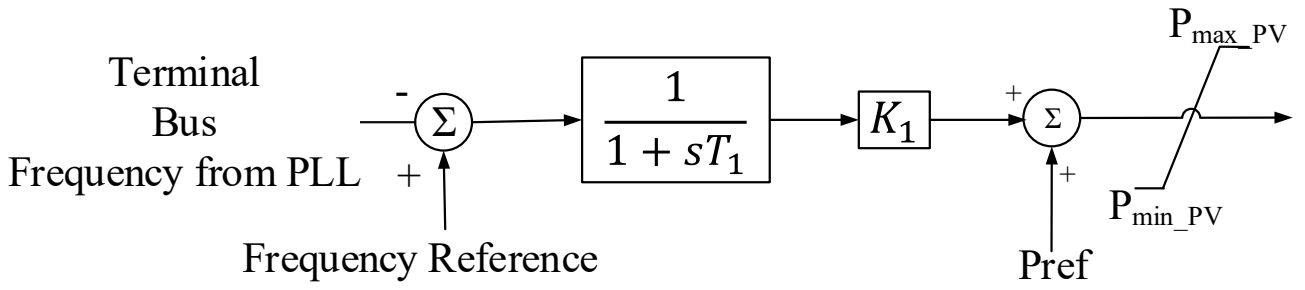

Fig. 12. Energy storage droop frequency control

2) Step response control: Step response control is usually adopted in remedial schemes and protection relays, for example, under-frequency load shedding. In step response, the contingency magnitude is estimated based on the rate of change of frequency (ROCOF) and system inertia. The step response magnitude is calculated by (2).

$$P_{step} = \alpha \cdot 2 \cdot H_{sys} \cdot \frac{ROCOF}{f_N} \cdot P_{sys} \tag{2}$$

where $\alpha$ is the ratio between the step response magnitude and the contingency magnitude. $H_{sys}$ is the system per unit inertia constant. $f_N$ is the nominal frequency. $P_{sys}$ is the system total load. After detecting a high ROCOF value, a time delay is applied before the frequency controller generating a step signal to the battery converter controller.

Fig. 11 shows the system frequency for the droop frequency control and step response control using battery energy storage. It can be seen that both control strategies can support primary frequency response, while the step response of batteries can arrest the frequency decline more quickly compared with droop frequency control. The reason is that step response can take advantage of the fast response characteristics of inverters and increase output very shortly after contingencies, thus helping to arrest frequency decline at an earlier stage. The step response control relies on the accurate estimation of the contingency magnitude and therefore its performance is sensitive to the accuracy of system inertia, load, and ROCOF values. A large error in contingency magnitude estimation may lead to insufficient or excessive response of energy storage.

### *E. Tactic ES2 — Energy Storage: Supercapacitor*

The high discharge current and long charge-discharge cycle life of a supercapacitor make it ideal for providing pulse power for primary frequency response. However, despite of high power density, supercapacitors can not provide sustained energy support for primary frequency response. Fig. 13 shows the frequency response of the droop control and step response control using supercapacitors in the 80% PV penetration scenario. It can be seen that the frequency experienced a second dip after the withdrawal of frequency support due to energy exhausting. Moreover, the frequency nadir of the second dip is slightly higher than the base case without supercapacitors due to the time constants of synchronous units' governor response. One major benefit of supercapacitors is delaying the frequency nadir and allowing other resources to response to frequency deviation. Compared with droop frequency control, step response control has stronger initial support but shorter duration, resulting in a lower frequency nadir.

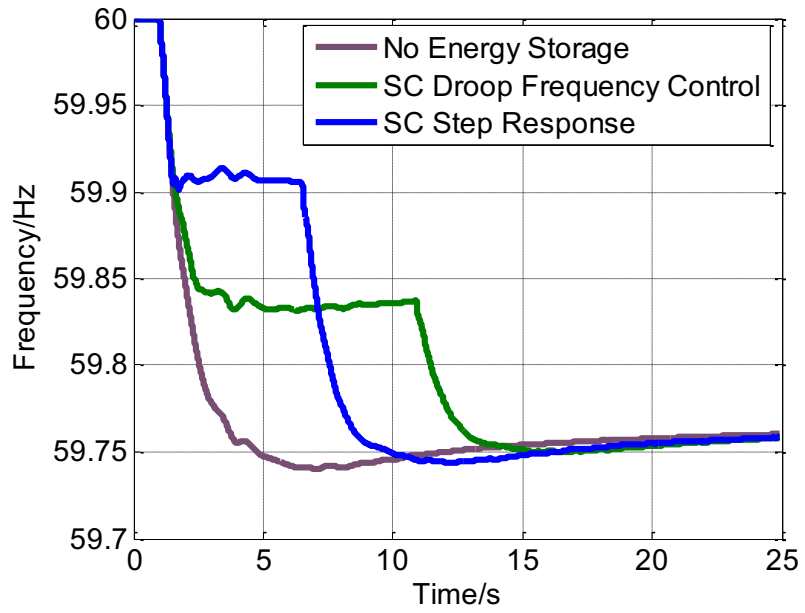

Fig. 13. EI frequency responses with droop frequency control and step response with supercapacitors (4.5 GW loss, 80% renewable)

In addition, it can be noticed from Fig. 13 that the EI primary frequency response is very sensitive to the timing of this short-term energy support from supercapacitors. To study this sensitivity, Fig. 14 and Fig. 15 show the change of EI frequency nadir and nadir time with different discharge time duration and energy capacities of supercapacitors in step response control. It can be seen that for a fix amount of supercapacitor energy, the system frequency nadir has a peak point for a certain time duration of discharge. This point is a balance between the first frequency dip (i.e. the initial frequency drop with frequency support from supercapacitors) and the second frequency dip (caused by the withdrawal of supercapacitors' output). Moreover, it can be noted that EI nadirs are within the range of 59.740Hz to 59.765Hz, which indicates that the nadir value in EI in not sensitive to both the energy amount and the discharge duration of supercapacitors. This is primarily due to the lazy 'L' characteristic (caused by the large system capacity and inertia) in the EI.

As the discharge duration prolongs, the first nadir becomes lower and second nadir becomes higher. Therefore, the overall frequency nadir time has a transition from the second nadir time to the first nadir time, which is significantly smaller, as shown in Fig. 15. Therefore, supercapacitors with higher energy can effectively delay the nadir and provide more time for secondary frequency response resources kicking in.

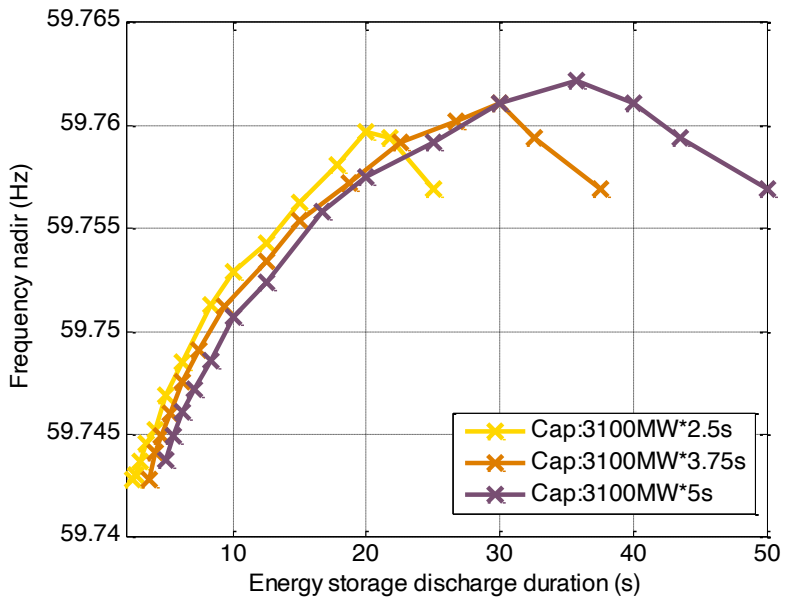

Fig. 14. EI frequency nadir change with discharge duration of supercapacitors (80% renewable)

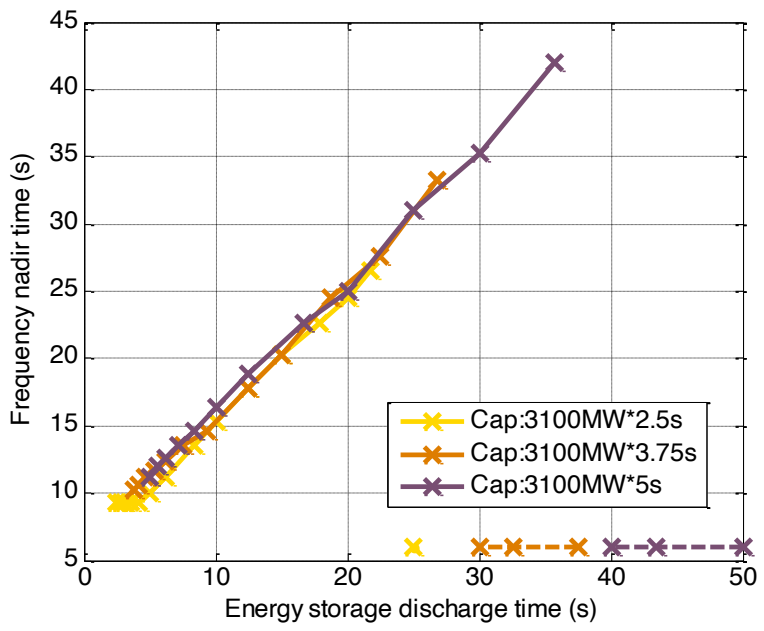

Fig. 15. EI frequency nadir time change with discharge duration of supercapacitors (80% renewable)

## IV. ERCOT System Frequency Response Enhancement Without Curtailing Solar Output

As noticed previously in Fig. 6, if no mitigation tactic is applied, the ERCOT system frequency will cross the under-frequency load shedding threshold when the renewable penetration rate reaches higher than 40%.

### *A. Tactic SG1 — Synchronous Generation: Reduce Governor Droop*

As a small system that is more concerned about primary frequency response, ERCOT requires that all available governors stay in service. Therefore, the governor ratio of ERCOT cannot be further increased unless more governor-controlled synchronous generators are committed to the grid [15]. In addition, as required by NERC, governor deadband across ERCOT had already been reduced to 16.7 mHz in 2014 [15]. Therefore, from the generation side, only the governor droop was adjusted to improve the ERCOT frequency response.

As shown in Fig. 16, reducing the governor droop from 5% (the current setting) to 3% will increase the ERCOT frequency nadir and settling frequency dramatically for the 20% scenario. However, different from the EI, adjusting governor droop was not able to substantially improve the ERCOT frequency response for the 80% renewable scenario as the number of synchronous units' governors was very limited. This result indicates that adjusting droop rates can hardly improve frequency response when the total capacity of governor-responsive generators is small and the reserve is too concentrated to a few generators. Therefore, for a system that is substantially deficient in frequency support, increasing the total capacity and number of governor-responsive generators is critical to enhance frequency stability.

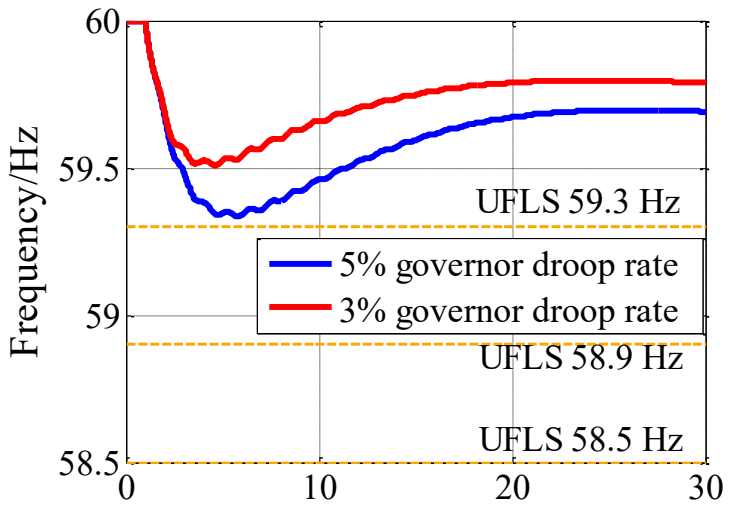

(a) 20% renewable

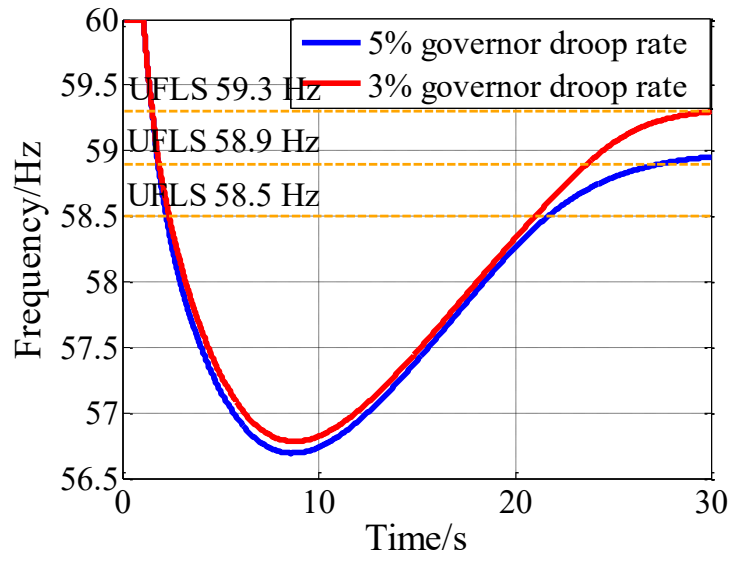

(b) 80% renewable

Fig. 16. ERCOT frequency responses with different governor droops (2.75 GW generation loss)

## *B. Tactic FRL — Load: Procure Fast Responsive Load*

Since governor droop adjustment is ineffective under high PV penetration due to the limited number of governor-responsive units, EROCT has begun to leverage fast responsive load for primary frequency response. In this study case, 2.5 GW fast responsive load is available to response to frequency decline. The load trips after a 0.5s delay as the frequency crossing 59.7 Hz. The ERCOT frequency response in Fig. 17 demonstrates significant increases in both the frequency nadir and the settling frequency after fast load response. It shows that fast responsive load is very effective for primary frequency response when the governor response provided by synchronous generators is insufficient.

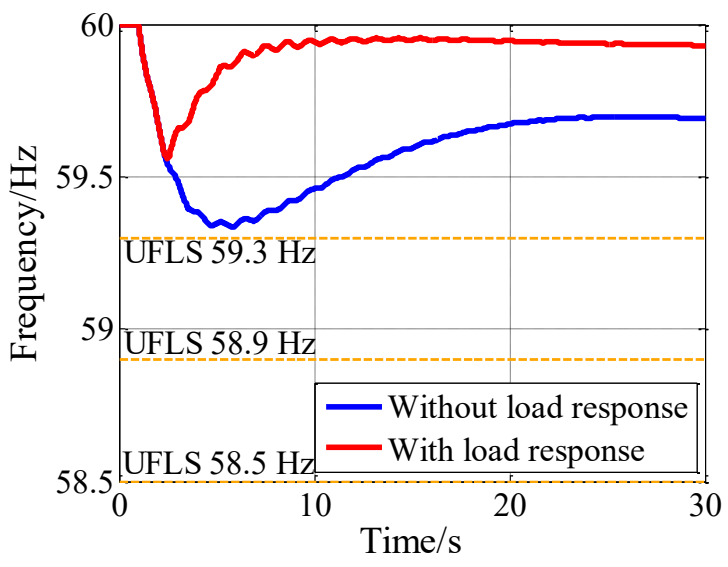

(a) 20% renewable

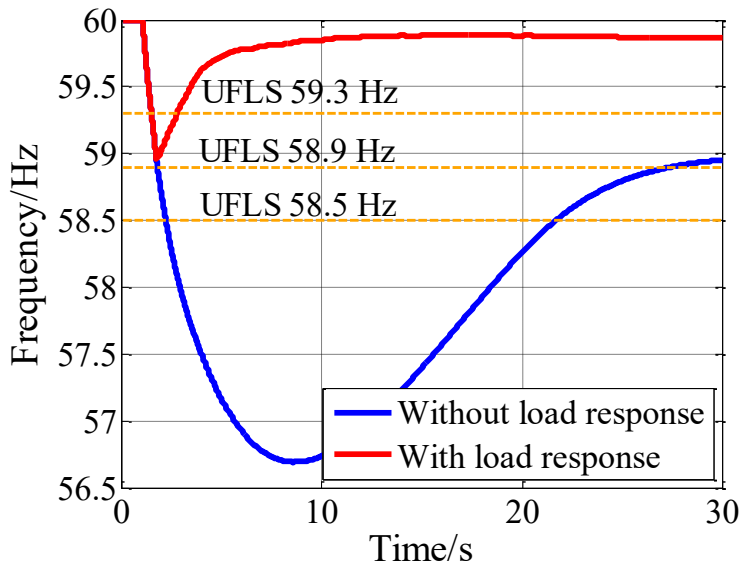

(b) 80% renewable

Fig. 17. ERCOT frequency responses with fast load response (2.75 GW generation loss)

## *C. Tactic ES1 — Energy Storage: Battery*

Battery energy storage for ERCOT frequency support applies the same controller as shown in Fig. 10. Control parameters are shown in Table 6. Fig. 18 shows the ERCOT frequency response with droop frequency control and step response of battery energy storage. It can be seen that the frequency is arrested at around 59.5 Hz and prevented from crossing UFLS thresholds. The frequency nadir values using the two control strategies are close. Compared with droop frequency control, step response control results in a higher settling frequency because of its larger response magnitude.

Table 6. Parameters of energy storage frequency control in the ERCOT

| Parameter | Value |
|---|---|
| Energy storage maximum output | 2,630MW |
| Number of energy storage installation locations | 50 |
| Droop frequency control droop rate | 5% |
| Droop frequency control deadband | 17 mHz |
| Step response threshold | 59.55 Hz |
| Step response delay for fault ride through | 0.5 s |
| Step response ratio α | 0.85 |
| Energy limit (default value for supercapacitors) | 2,630 MW*10s |

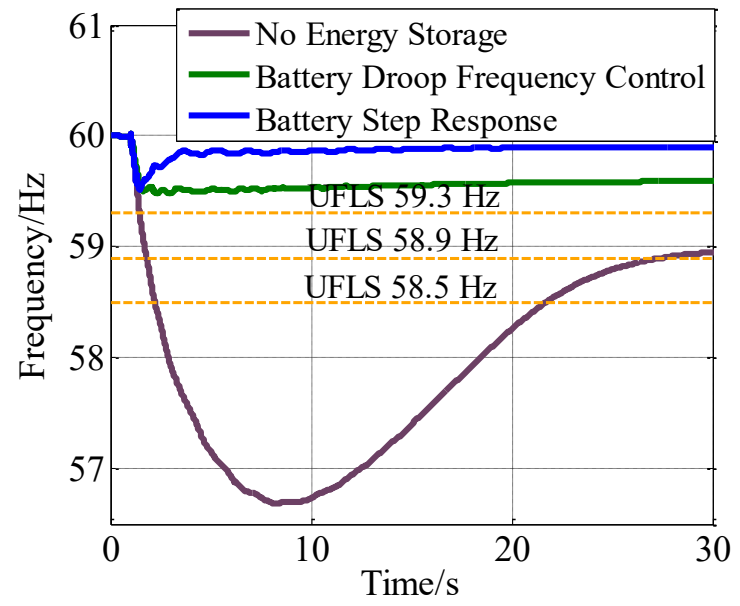

Fig. 18. ERCOT frequency responses using droop frequency control and step response control of batteries (2.75 GW loss, 80% renewable)

## *D. Tactic ES2 — Energy Storage: Supercapacitor*

In contrast to batteries, high-power-density supercapacitors were applied to the ERCOT system for primary frequency response. Fig. 19 shows the frequency regulation performance of supercapacitors using droop frequency control and step response control. Similar to the EI, withdrawal of supercapacitor causes frequency crossing the UFLS threshold at a later time. This delay in UFLS wins time for other resources to response. Compared with droop control, the frequency of step response control has a lower nadir and earlier nadir time. This is because step response control has larger output at the early stage, leading to earlier response withdrawal. Due to the time constants in governor response and consequent response delay, the system has a lower nadir.

Fig.18 and Fig. 19 show the change of frequency nadir and nadir time with different discharge time duration and energy capacities of supercapacitors. It can be seen that the sensitivity of ERCOT frequency response has a similar pattern to that of the EI: releasing energy as fast as possible does not guarantee the highest frequency nadir because of earlier output withdrawal. Moreover, compared with EI, varying discharge duration has relatively larger impact on the nadir in ERCOT. It can be noticed that the maximum nadir under high PV penetration can hardly be increased by either increasing the capacity or the maximum power of supercapacitors. Consistent with the EI, this is because a larger energy capacity of supercapacitors can only delay the time of the frequency nadir but can hardly increase the nadir value. This result also indicates that achieving the best frequency regulation effects using supercapacitors will require coordination with other frequency response resources.

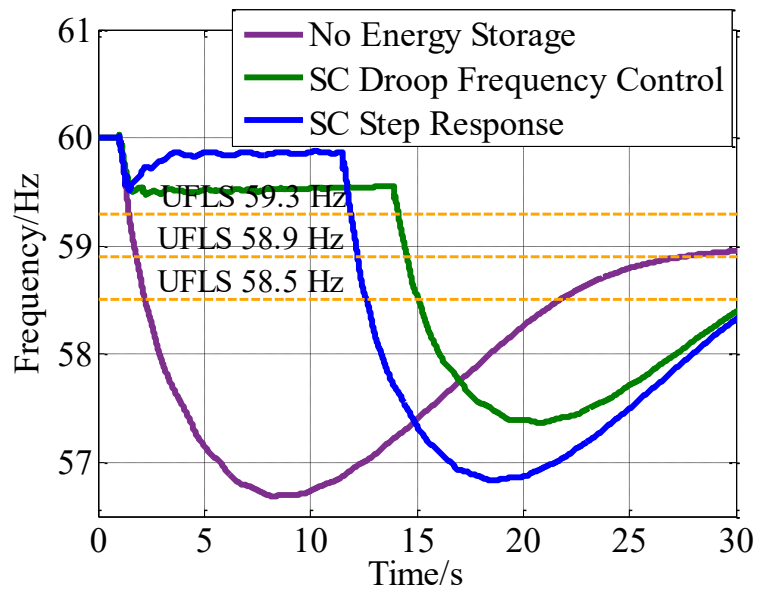

Fig. 19. ERCOT frequency responses with droop frequency control and step response using supercapacitors (80% renewable)

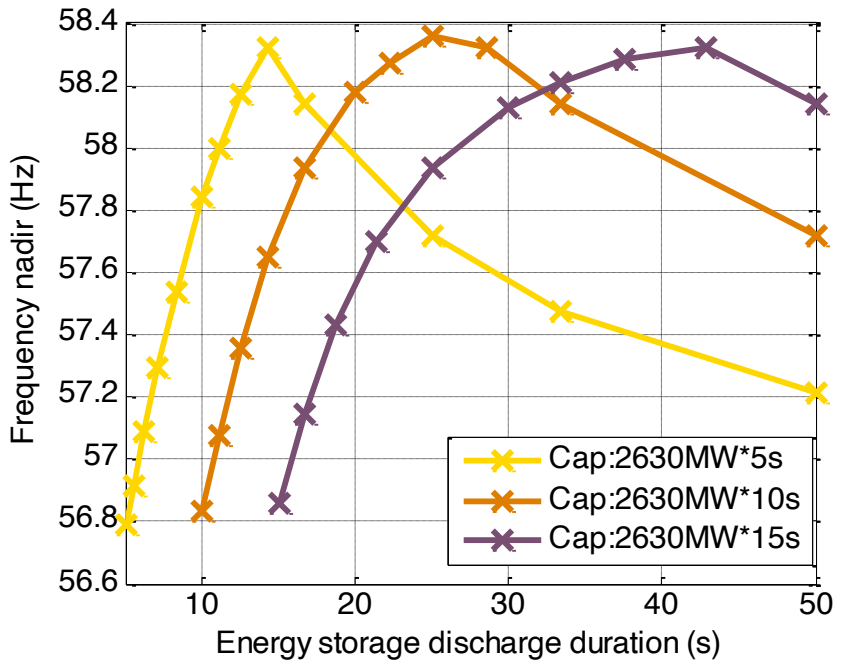

Fig. 20. ERCOT frequency nadir change with discharge duration of supercapacitors (80% renewable)

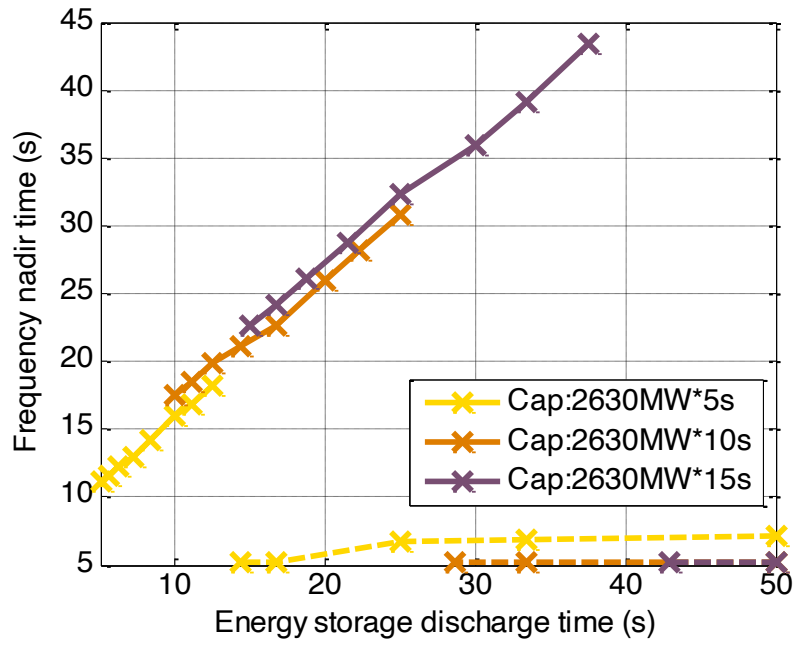

Fig. 21. ERCOT frequency nadir time change with discharge duration of supercapacitors (80% renewable)

## V. Comparison of Mitigation Tactics in the Two Systems

A comparison of the performance of various tactics is summarized in Fig. 22, which is further explained as follows:

1. *Tactic SG1 — Decrease Governor Droop Rates; and SG2 — Decrease Governor Deadband:* Adjusting governor droop rates and deadbands do not require additional equipment, so they can be conveniently implemented. However, their effectiveness is limited in high renewable penetration scenarios: decreasing governor droop rates is not effective in improving the frequency nadir when governor response resources are insufficient, and changing governor deadbands can only improve frequency response noticeably when frequency deviations are small. For example, as shown in the case study, decreasing governor droop in the ERCOT 80% renewable scenario is far less effective than the same tactic in the EI. In addition, decreasing governor droop rates requires additional headroom and adds generation opportunity costs; and a very small droop rate may lead to instability. In addition, too small deadbands will increase wear-out of governing systems.

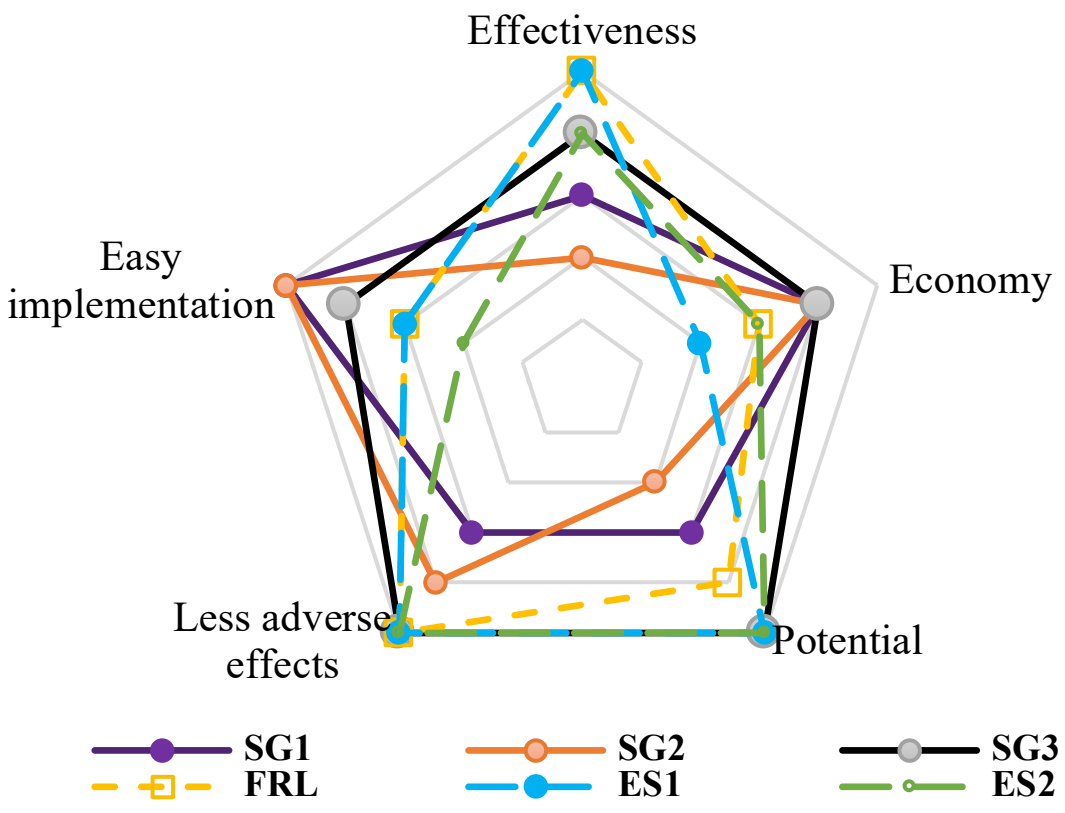

Fig. 22. High-level comparison on the attributes of different tactics

2. *Tactic SG3 — Increase Governor Ratio:* Increasing governor ratios can improve primary frequency response effectively in a range of renewable penetration scenarios. It may require considering additional constraints in unit commitment and installation of new governor systems on some generators.
3. *Tactic FRL — Procure Fast Responsive Load:* Fast responsive load is one of the most effective approaches, especially when frequency response capabilities on the generation side have been exhausted in high renewable penetration scenarios. As shown in the case study, for both the EI and the ERCOT, fast responsive load can provide robust frequency support, and its effectiveness is insensitive to system characteristics. Its disadvantage is the associated costs of load control services, which should be specified by agreements between large industrial customers and system operators.
4. *Tactic ES1 — Battery Energy Storage:* Battery energy storage can arrest frequency decline effectively using frequency droop control or step-response control.

Compared with frequency droop control, the step-response control is more effective in arresting frequency decline if the power imbalance estimation has enough accuracy. As the price of batteries continues decreasing, it will become more economically competitive for primary frequency response, especially for applying in smaller power grids.

5. *Tactic ES2 — Supercapacitor Energy Storage:* As supercapacitors energy storage can discharge only for a short term to support the grid frequency, there exists a certain discharge duration that can maximize the frequency nadir, corresponding to a balance between the initial frequency dip caused by the contingency and secondary frequency dip caused by energy exhausting. This maximum nadir value does not change much when varying the energy capacity or the rated output power of supercapacitors due to their short-term support characteristic. Compared with that of small systems (e.g. the ERCOT system), the maximum nadir value of a larger interconnection grid (e.g. the EI system) is further less sensitive to the supercapacitor discharge duration due to large inertia and unobvious frequency recovery.

## VI. Conclusions

To improve primary frequency response without curtailing solar energy in high PV interconnections, this paper studied the performance of various tactics in the U.S. EI and ERCOT interconnection grids. A comparison study between various mitigation tactics provides meaningful information for power system planners, operators, reliability coordinators, and regulators on how to utilize available resources to ensure primary frequency response capabilities in PV penetration. As the two interconnection grids have different frequency response features in their current and future high-renewable scenarios, the result comparison between the two systems and generalized conclusions provide reference value for stakeholders of high renewable power systems in general.